%% file: ieee_wcm_main.tex
\documentclass[journal]{IEEEtran}
\usepackage[printonlyused]{acronym}
\usepackage[style=ieee,maxbibnames=99]{biblatex}
\bibliography{biblio}
\usepackage{amsmath,amsfonts}
\usepackage{algorithmic}
\usepackage{algorithm}
\usepackage{array}
\usepackage{textcomp}
\usepackage{stfloats}
\usepackage{url}
\usepackage{verbatim}
\usepackage{graphicx}
\usepackage{xcolor}
\usepackage{tabularray}
\usepackage{balance}
\usepackage{booktabs}
\usepackage{enumitem}
\usepackage{subfigure}

\begin{document}

\title{Unified Evaluation Methodology for AI-Native Integrated Sensing and Communication}

\author{Filip Lemic, Andra Blaga, Francesco Devoti, Guillermo Encinas Lago, Jan Adler, Amitha Mayya, Padmanava Sen, Giorgos Stratidakis, Sotiris Droulias, Angeliki Alexiou, Alexander Artemenko, Aya Mostafa Ahmed, Visa Koivunen, Robin Rajamäki, Simon Schütze, Robert Elschner, Amélie Hennequart, Ahmad Shoukair, Youssef Nasser, Nahuel Soprano-Loto, François Baccelli, Visa Tapio, Paul Almasan, Andra Lutu, Vincenzo Sciancalepore, Carmen Delgado, Xavier Costa-Pérez
\thanks{F. Lemic (email: filip.lemic@i2cat.net), A. Blaga, G. Encinas, C. Delgado, and X. Costa-Pérez are with i2CAT Foundation, Spain. F. Devoti, V. Sciancalepore, and X. Costa-Pérez are with NEC Labs Europe GmbH, Germany. X. Costa-Pérez is also with ICREA, Spain. J. Adler, A. Mayya, and P. Sen are with Barkhausen Institute, Germany. G. Stratidakis, S. Droulias, and A. Alexiou are with University of Piraeus, Greece. A. Artemenko and A. Ahmed are with Robert Bosch GmbH, Germany. V. Koivunen is with Aalto University, Finland. R. Rajamäki is with Tampere University, Finland. S. Schütze and R. Elschner are with Fraunhofer HHI, Germany. A. Hennequart, A. Shoukair, and Y. Nasser are with Greenerwave, France. N. Soprano-Loto and F. Baccelli are with Inria Paris, France. V. Tapio is with University of Oulu, Finland. P. Almasan and A. Lutu are with Telefónica I\&D, Spain.}
\vspace{-3mm}}%

\markboth{IEEE DRAFT}%
{Shell \MakeLowercase{\textit{et al.}}: A Sample Article Using IEEEtran.cls for IEEE Journals}

\maketitle

\begin{abstract}
\ac{ISAC} couples radio sensing, data transmission, and control actions within a single closed-loop system. When \ac{AI}-driven policies adapt sensing and communication online across a variety of sensing tasks and objectives, end-to-end performance is shaped not only by waveform and channel conditions but also by inference latency, uncertainty, environmental dynamics, and hardware non-idealities, leading to fundamental trade-offs between sensing accuracy, communication reliability, and resource overhead. This manuscript presents a unified system architecture and evaluation methodology for \ac{AI}-native \ac{ISAC}, defined as \ac{ISAC} in which learning-based agents adapt sensing, communication, and actuation policies online under uncertainty. We formalize the design space of closed-loop \ac{ISAC}, propose a three-stage validation pipeline from bounds and feasibility analysis, through high-fidelity digital-twin simulation, to preliminary over-the-air validation, and provide a minimal reporting checklist that links technical \acp{KPI} (e.g., data rate, SINR, target detection, parameter estimation, track quality, localization error, outage, latency, overhead, and energy per decision) to application-level \acp{KVI} (e.g., availability and mission effectiveness). Two representative instantiations, specifically \ac{UAV}-based outdoor and \ac{RIS}-enabled indoor coverage extensions, are used to illustrate how to structure reproducible baselines and comparable evidence across heterogeneous deployments, helping bridge the gap between theoretical \ac{ISAC} gains and deployment-ready performance claims.
\end{abstract}


\input{acronym_def}
\input{introduction}

\input{system_architecture}
\input{evaluation_methodology}
\input{scenarios}
\input{results}
\input{conclusion}

\section*{Acknowledgments}
This work was supported by the European Union's Horizon Europe's programme (grant 101139161 - INSTINCT project). 

\renewcommand{\bibfont}{\footnotesize}
\printbibliography





\end{document}

%% file: acronym_def.tex

\acrodef{5G}[5G]{Fifth Generation}
\acrodef{AI}[AI]{Artificial Intelligence}
\acrodef{AoA}[AoA]{Angle-of-Arrival}
\acrodef{DoA}[AoA]{Direction-of-Arrival}
\acrodef{POMDPs}{Partially Observable Markov Decision Processes}
\acrodef{ASR}[ASR]{Angular Spectrum Representation}
\acrodef{BER}[BER]{Bit Error Rate}
\acrodef{BS}[BS]{Base Station}
\acrodef{CRB}[CRB]{Cramér-Rao Bound}
\acrodef{CSI}[CSI]{Channel State Information}
\acrodef{CFAR}{Constant False Alarm Rate}
\acrodef{DSP}[DSP]{Digital Signal Processing}
\acrodef{FI}[FI]{Fisher Information}
\acrodef{GNSS}[GNSS]{Global Navigation Satellite System}
\acrodef{I3}[I3]{Interactive, Immersive, and Intelligent}
\acrodef{IoT}[IoT]{Internet of Things}
\acrodef{ISAC}[ISAC]{Integrated Sensing and Communication}
\acrodef{KPI}[KPI]{Key Performance Indicator}
\acrodef{KVI}[KVI]{Key Value Indicator}
\acrodef{LoS}[LoS]{Line-of-Sight}
\acrodef{ML}[ML]{Machine Learning}
\acrodef{MI}{Mutual Information}
\acrodef{MSE}[MSE]{Mean Squared Error}
\acrodef{NLoS}[NLoS]{Non-Line-of-Sight}
\acrodef{PEB}{Position Error Bound}
\acrodef{RIS}[RIS]{Reconfigurable Intelligent Surface}
\acrodef{RSRP}[RSRP]{Reference Signal Received Power}
\acrodef{RCS}{Radar Cross Section}
\acrodef{SNR}[SNR]{Signal-to-Noise Ratio}
\acrodef{SRS}[SRS]{Sounding Reference Signal}
\acrodef{ToA}[ToA]{Time-of-Arrival}
\acrodef{ToF}[ToF]{Time-of-Flight}
\acrodef{UAV}[UAV]{Unmanned Aerial Vehicle}
\acrodef{UE}[UE]{User Equipment}

%% file: introduction.tex
\vspace{-1mm}
\section{Introduction}

The convergence of native radio sensing, next-generation wireless connectivity, and data-driven decision making enables a new class of \ac{I3} systems~\cite{Droulias2026I3ISAC} in which the network supports both connectivity and environment-aware functions. In such systems, radio infrastructure must provide throughput, coverage, localization, mapping, and sensing-assisted control under strict latency and reliability constraints. \acf{ISAC} enables this integration by allowing radio signals, spectrum, and hardware to be shared across sensing and communication tasks~\cite{Droulias2026I3ISAC}. \ac{ISAC} systems couple measurement quality, resource allocation, and actuation decisions (e.g., platform mobility or environment programmability) in a single closed-loop system. In practice, realized gains are sensitive to the dynamics of radio environments, interference variability, hardware non-idealities, and real-time constraints such as limited energy budgets and processing delays \cite{islam2024mobile}. Coverage extension is a representative use case~\cite{blaga20243dsar,encinas2025riloco,encinas2025coloris} because it forces operation under partial observability (e.g., due to sensor noise, environmental state uncertainty, blockage, and \ac{NLoS} conditions) and tight resource constraints, making the sensing, communication, and actuation coupling explicit.

A key source of complexity is the increasing use of learning-based policies to operate \ac{ISAC} in a closed-loop manner. In \emph{\acf{AI}-native \ac{ISAC}}, learning-based agents adapt sensing, communication, and actuation policies online under uncertainty (e.g., sensing overhead, scheduling, beam configuration, or platform mobility). A key enabler of \ac{AI}-native \ac{ISAC} operating in closed-loop mode is situational awareness about the operational environment. This differs from traditional evaluations that often assess waveform or link-level performance in isolation and assume static or single open-loop configurations. In learned closed-loop control, end-to-end behavior becomes sensitive to inference latency, stochastic actions (e.g., by other agents), distribution shift (i.e., rapid changes in underlying probability models), and implementation constraints. Consequently, single protocol layer metrics no longer predict system-level performance. These restricted evaluation strategies under-represent how sensing uncertainty propagates into resource allocation and actuation. Furthermore, they overlook how overhead and scheduling constrain sensing updates, and how real-time inference affects stability, safety, and availability. As a result, performance claims may be difficult to reproduce and may not transfer across environments or hardware platforms.

Although the \ac{ISAC} paradigm is foundational for next-generation networks, the wireless community lacks a standardized evaluation methodology. Just as unified methodologies were important for transitioning communication (e.g., cognitive radio~\cite{zhao2009performance}), sensing (e.g., localization~\cite{van2015platform}), and \ac{AI}-based network orchestration from research to practice, they are equally needed for \ac{AI}-native \ac{ISAC}. Radar sensing has its well-defined \acfp{KPI} stemming from detection theory, parameter estimation, tracking, optimal filtering, control, and phenomena explaining how signals interact with targets and the environment. These unified evaluation methodologies established well-defined \acfp{KPI}, matched baselines, and consistent reporting conventions to enable fair comparisons between different approaches and solutions. We adapt these established practices in this work.

Specifically, we first formalize the closed-loop \ac{ISAC} design space by identifying the principal coupling pathways between sensing, communication, and \ac{AI}-driven control (e.g., reinforcement learning) in a unified system architecture. Building on this foundation, we propose a three-stage evaluation framework that provides realistic predictions of the performance while maintaining traceability: i) bounds and feasibility analysis under controlled assumptions, ii) high-fidelity digital twin simulation for end-to-end trade-off exploration, and iii) experimental validation via emulation and/or over-the-air measurements. The framework is complemented by a minimal reporting checklist that makes assumptions, baselines, and \ac{KPI} to \acf{KVI} mappings explicit, enabling repeatable benchmarking across heterogeneous deployments.

\begin{figure*}[!t]
\centering
\includegraphics[width=0.86\linewidth]{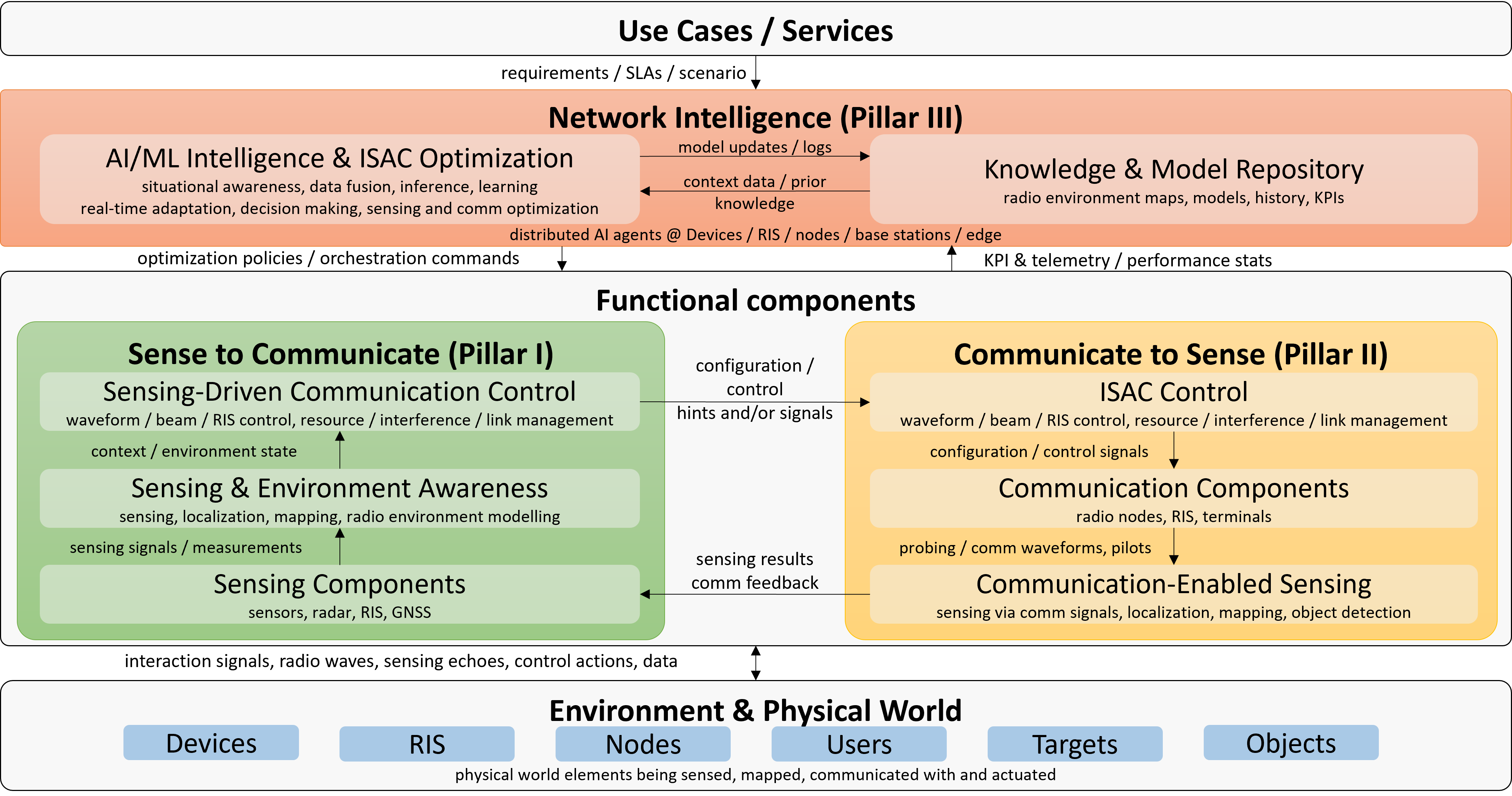}
\vspace{-1mm}
\caption{Reference \ac{AI}-native \ac{ISAC} system architecture, illustrating the interplay between the physical world, functional components, and the network intelligence.}
\vspace{-3mm}
\label{fig:system_architecture}
\end{figure*}

We use practically relevant scenarios to make the methodology actionable and to highlight the role of actuation in closed-loop \ac{ISAC}. First, \acfp{UAV}-assisted outdoor coverage extension illustrates mobility control, where repositioning changes measurement geometry and increases \ac{LoS} likelihood and coverage/signal quality in infrastructure-limited areas. Second, \acfp{RIS}-enabled indoor coverage extension illustrates environment programmability, where controllable reflections can improve the radio environment in occluded regions. These are presented as examples of practical mechanisms illustrating the end-to-end coupling needed in \ac{AI}-native \ac{ISAC} evaluation.

\begin{figure*}[!t]
\centering
\includegraphics[width=0.7\linewidth]{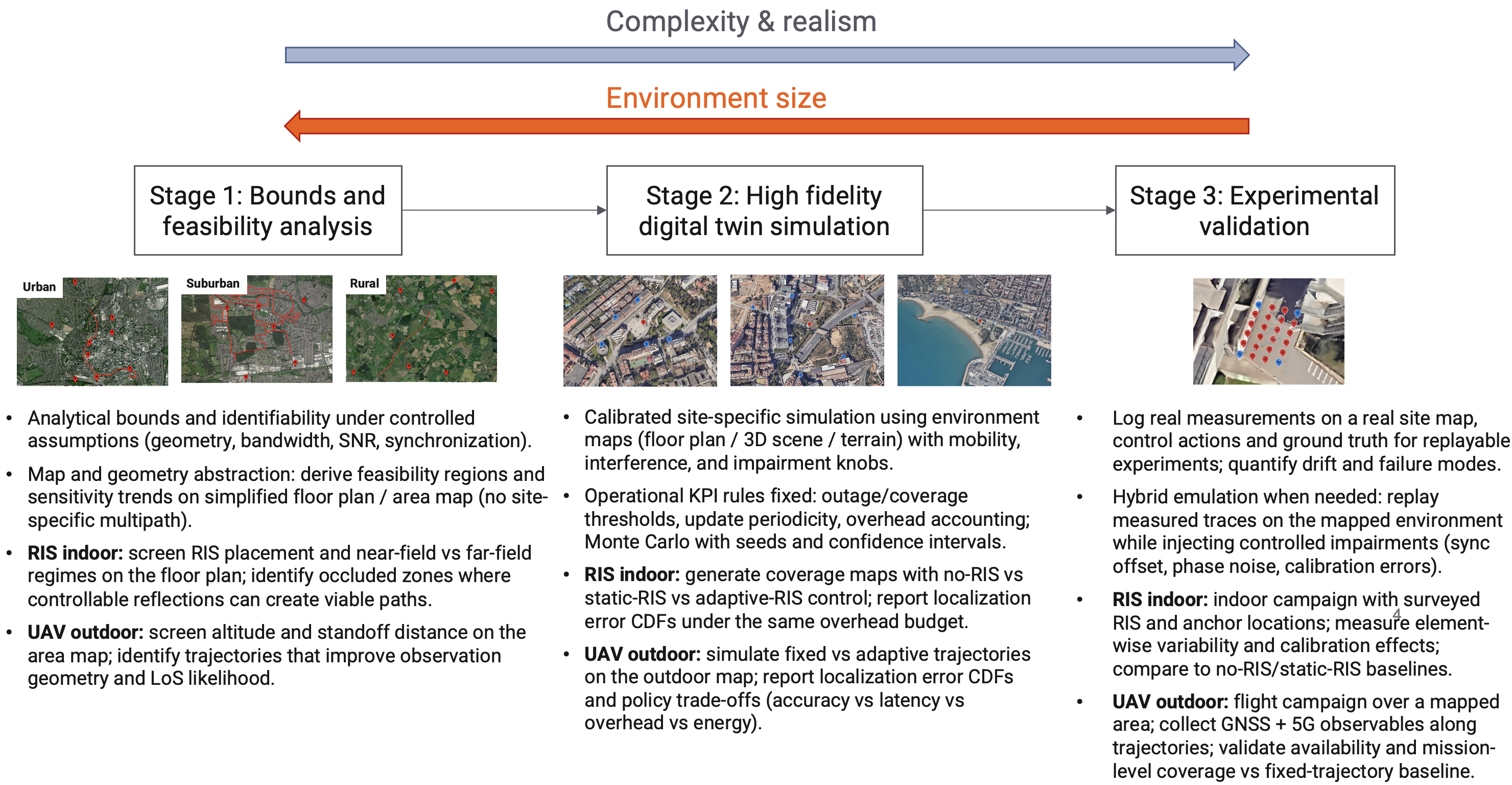}
\vspace{-1mm}
\caption{Three-stage evaluation methodology for \ac{AI}-native \ac{ISAC}. The framework progresses from bounds and feasibility analysis, through high-fidelity digital twin simulation, to experimental validation using emulation and over-the-air testbeds.}
\vspace{-3mm}
\label{fig:evaluation_methodology}
\end{figure*}

The main contributions are: i) We formalize the design space of closed-loop \ac{ISAC} by identifying key architectural elements and coupling pathways between sensing, communication, and \ac{AI}-driven control. ii) We propose a three-stage evaluation framework comprised of bounds and feasibility analysis, high-fidelity digital twin simulation, and experimental validation, with consistent assumptions and reporting across stages. iii) We define a minimal \ac{KPI} and \ac{KVI} reporting set and baseline taxonomy that supports repeatable benchmarking and comparability across heterogeneous deployments. iv) We provide two practically relevant example scenarios of coverage extension in \ac{RIS}-enabled indoor and \ac{UAV}-assisted outdoor deployments, to illustrate how to structure reproducible baselines and comparable evidence under practical constraints.

%% file: system_architecture.tex
\vspace{-1mm}
\section{System Architecture}

Aligned with the concept of \ac{ISAC} for future \ac{I3} connectivity~\cite{Droulias2026I3ISAC}, this work develops an operational evaluation methodology for \ac{ISAC}. To enable fair, comparable, and reproducible evaluation across heterogeneous deployments, we define a reference system architecture that models an \ac{ISAC} network as a coherent cyber-physical system connecting the physical world to a network intelligence layer. The architecture is deliberately use-case agnostic and supports coverage extension for sensing-enabled services across indoor scenarios via radio environment programming using \acp{RIS} and outdoor scenarios via mobility of \acp{UAV}. It integrates end-user applications, radio access infrastructure, sensing subsystems, and the respective components into a unified system model in which sensing outputs inform decisions that, in turn, modify the radio scene and subsequent observations.

As shown in Figure~\ref{fig:system_architecture}, the architecture is organized around three interacting functional pillars that make the closed-loop operation explicit with quantitative performance criteria~\cite{Droulias2026I3ISAC}. The first pillar, \textit{Sense to Communicate}, uses sensing observations (e.g., state of the radio environment, user localization, object mapping, and \ac{CSI}) to adapt communication parameters. By providing situational awareness to the network intelligence layer, the system can adjust beam management, scheduling, handover policies, and actuation states, and choose optimal actions based on the state of the operational environment. For example, when sensing indicates emerging blockage or a newly occluded region, the controller may trigger a beam switch or handover, or reconfigure a \ac{RIS} to improve coverage and observability, as in \ac{AI}-driven solutions such as \textit{RiLoCo}~\cite{encinas2025riloco}.

Conversely, the second pillar, \textit{Communicate to Sense}, repurposes communication waveforms and hardware for environmental perception~\cite{Droulias2026I3ISAC}. It governs the dual use of reference signals and pilots intended for synchronization and channel estimation to extract range, velocity, and angle information from reflections using, e.g., standard multicarrier waveforms~\cite{koivunen2024multicarrier}, enabling sensing without dedicated spectrum. This pillar enforces the trade-off between sensing and communication at the frame and scheduling levels by controlling pilot overhead and update intervals so that sensing performance improves without violating latency and reliability targets.

The third pillar, \textit{Network Intelligence}, governs these interactions by implementing closed-loop decision making under uncertainty~\cite{Droulias2026I3ISAC}. It hosts \ac{AI} agents that fuse sensing and communication streams and select actions such as frequency allocations, spatial domain beampatterns, task prioritization, and changing values of operational parameters (e.g., beam, actuation state, or trajectory). This layer explicitly accounts for sensing uncertainty, inference latency, control overhead, and actuation limits, enabling end-to-end evaluation in terms of both technical \acp{KPI} (e.g., target detection, target parameter estimation, track quality, localization error, outage, latency, overhead, and energy per decision) and application-level \acp{KVI} (e.g., availability and mission effectiveness).

\textit{Actuation mechanisms: programmable radio environments and mobility.}
A central architectural element of closed-loop \ac{ISAC} is actuation, meaning the selection of actions that change either the radio environment or the observation geometry. From a learning point of view, the action space is much larger, starting from waveform selection, power allocation, frequency allocation, to spatial allocation. We include programmable propagation control using \acp{RIS} as a representative technology, which can tune reflections and create favorable propagation conditions, including virtual \ac{LoS} paths in cluttered environments~\cite{direnzo2020smart}. Within our architecture, the \ac{RIS} is a controllable component of the loop that can be used to improve coverage and support sensing by increasing geometric diversity through controllable reflections~\cite{encinas2025riloco}.

The same abstraction accommodates mobility-driven coverage extension, where a mobile \ac{BS} (e.g., a \ac{UAV}) serves as a controllable \ac{ISAC} node. By adapting its trajectory, the system can improve \ac{LoS} probability and make the measurement geometry more favorable, trading off accuracy against overhead, latency, and energy. This parallel between \ac{RIS}-based environment programmability and \ac{UAV}-supported mobility control is central to our evaluation approach, as it exposes the closed-loop coupling for \ac{AI}-native \ac{ISAC} management.

%% file: evaluation_methodology.tex
\section{Multi-Stage Evaluation Methodology}
Evaluating \ac{AI}-native \ac{ISAC} as a coupled closed-loop system requires a unified methodology that provides realistic predictions of performance while preserving traceability and repeatability. 
System architectural assumptions, baseline definitions, and reporting conventions must be derived from analytical performance bounds and physical constraints related to target interaction and radio propagation. These foundations must then be validated in simulations and carried over to experimental validation, so that measured gains can be attributed to concrete design choices rather than scenario-specific artifacts.

This is particularly important for coverage extension as a representative use case, where improvements can arise from different actuation mechanisms such as environment programmability (e.g., \ac{RIS}) or mobility control (e.g., \ac{UAV}) and must be compared under consistent conditions. We therefore adopt a three-stage evaluation approach that provides (i) early insight via feasibility checks and scaling trends, (ii) broad exploration of architectural and algorithmic options under controlled variability, and (iii) confirmation under real-world impairments through experimental validation (Figure~\ref{fig:evaluation_methodology}).
To make studies comparable across deployments, the evaluation pipeline must standardize reporting. Table~\ref{tab:protocol_checklist} integrates a minimal stage-specific protocol with a compact set of \acp{KPI} and \acp{KVI}, including their definitions, scenario instantiation anchors, and representative target benchmarks.

\begin{table*}[!t]
\caption{Minimal evaluation protocol and reporting checklist for reproducible \ac{AI}-native \ac{ISAC} assessment, integrating primary \acp{KPI} and \acp{KVI} with operationalization, instantiation anchors, and representative targets.}
\label{tab:protocol_checklist}
\vspace{-1mm}
\centering
\footnotesize
\renewcommand{\arraystretch}{1.08}
\setlength{\tabcolsep}{4pt}
\begin{tabular}{|p{1.0cm}|p{5.2cm}|p{5.4cm}|p{5.4cm}|}
\hline
\textbf{Stage} & \textbf{Inputs (must specify)} & \textbf{Outputs (must report)} & \textbf{Baselines and statistics} \\
\hline
\textbf{Stage 1: Bounds} &
Geometry (anchors, \ac{RIS} placement, aperture size), carrier, bandwidth, synchronization, SNR (incorporating antenna gains, target \ac{RCS}), and noise &
CRB or \ac{PEB} (or \ac{ToA} and \ac{AoA} bounds), coverage probability and rate trends, feasibility regions (identifiability) &
Baselines: no \ac{RIS}, far field versus near field, ideal synchronization versus offset. Report assumptions explicitly (units, priors). \\
\hline
\textbf{Stage 2: Digital twin} &
Site model (maps), mobility model, traffic and interference model, algorithm settings (codebooks, update rate), impairment parameters, operational KPI rules (e.g., outage definition) &
Coverage maps and CDFs; distributions (CDF and percentiles) of localization error, latency, overhead; trade-offs between rate, error, latency, and energy &
Baseline: static beams, non-adaptive control, communication only vs. sensing assisted. MC: number of runs, seeds, confidence intervals; sweep ranges for impairment parameters. \\
\hline
\textbf{Stage 3: Experimental} &
Hardware chain, clocking/sync strategy, calibration, ground truth method and accuracy, logging schema, geometry, probing settings &
End-to-end KPIs with uncertainty, robustness to drift, failure mode characterization, and reproducible traces enabling replay/emulation &
Baseline: identical hardware, \ac{AI} disabled/fixed policy. Report repeats, confidence intervals, impairment sensitivity (sync offsets, drift). \\
\hline

\multicolumn{4}{|p{\dimexpr\textwidth-2\tabcolsep-2\arrayrulewidth\relax}|}{
\footnotesize
\begin{tabular}{@{}p{2.7cm}|p{\dimexpr\textwidth-2\tabcolsep-2\arrayrulewidth-2.9cm\relax}@{}}

\textbf{Primary indicators} &
\textbf{Sensing KPIs:} Localization accuracy as 2D or 3D error reported via RMSE and CDF percentiles (e.g., median and 80th percentile) versus ground truth, with confidence intervals.
Target detection performance measured via detection probability at a specified false alarm rate (e.g., \ac{CFAR}).
Range resolution (waveform and bandwidth dependent) and angle resolution (aperture size dependent) as minimum distinguishable target separations, validated in controlled tests.

\textbf{Communication KPIs:} \ac{BER} measured under operational conditions using known transmit and receive sequences.
Coverage and outage via spatial \ac{RSRP} maps and a declared outage rule (example rule \ac{RSRP} below $-100$~dBm), with coverage extension reported relative to a baseline.

\textbf{Joint KPIs:} Latency as end-to-end delay from sensing to inference and control to actuation and feedback, including processing and transmission, reported with mean and a tail percentile (e.g., 95th), and a breakdown when available.
Overhead as the fraction of resources used for sensing and control (pilots, feedback, reconfiguration signaling) together with update periodicity, with baselines matched in overhead (example pilot density UL \ac{SRS} at 100~Hz).

\textbf{Value KVIs:} Positioning availability as the fraction of time or locations meeting the stated accuracy and latency constraints, reported for the mission or site.
Coverage extension is the relative increase in area meeting KPI thresholds versus a baseline (representative target benchmark: two times coverage in occluded areas).
Operational efficiency as a task-level proxy, such as reduced downtime attributed to stable control (representative benchmark: more than 20 percent reduction).
Safety improvement as a proxy linked to accuracy and reaction time, such as missed detection rate (representative benchmark: less than 1 percent false negatives).
Energy efficiency as energy per mission or per update/decision, including compute and signaling attribution. \\ \hline

\textbf{Instantiation anchors} &
\textbf{Indoor:} LiDAR ground truth at cm level and multi-band evaluation (e.g., 2, 3.5, 10~GHz) with no \ac{RIS} and static \ac{RIS} baselines. \newline
\textbf{Outdoor:} surveyed grid or reference transfer for ground truth, representative evaluation geometry (e.g., $10\times 6$~m area, four gNB corners, 20 UE points on a 2~m grid), and representative \ac{5G} settings (n78, 30~kHz subcarrier spacing, up to 40~MHz bandwidth), with a fixed scan trajectory as baseline. \\ \hline

\textbf{Cross stage reporting} &
Explicit KPI to KVI mapping (e.g., availability as the fraction of locations where error and latency thresholds are met), explicit thresholding (outage and accuracy rules), matched overhead accounting across baselines, and energy per update or decision. \\
\end{tabular}
} \\
\hline
\end{tabular}
\vspace{-3mm}
\end{table*}

\noindent\textbf{Stage 1: Bounds and feasibility analysis.}
Stage 1 provides fast and interpretable reference points that are easy to compute and scale across parameter sweeps. It abstracts site-specific and hardware details to answer two questions early: is the problem identifiable under the assumed sensing modality, and which parameters dominate performance trends. 
We use estimation theoretic bounds such as the \ac{CRB} to quantify the best achievable sensing error variance under controlled assumptions for tasks such as \ac{ToA} and \ac{AoA} estimation. Additionally, information theoretic metrics such as \ac{MI} are commonly used in this context, maximizing MI related to acquiring maximally information from the target and in communication rate, which has a direct connection to the \ac{MSE}. For target detection, fundamental limits regarding the probability of detection under probability of false alarm constraints serve as further reference points. For ToA, the CRB depends on the bandwidth and SNR, while for \ac{AoA}, resolution depends on the aperture size, which is important if you operate in multi-target scenarios, perform beamforming, or \ac{DoA} estimation. To capture coverage and interference trends at
network scale, we complement these with stochastic geometry models
that yield tractable expressions for coverage probability and
Shannon rate in dense deployments. 
This foundational stage can broadly accommodate evaluations of track quality, for example via \ac{MSE}, or track sharpness considering the main axis of error ellipsoids.
For \ac{RIS}-assisted propagation, we additionally use a wave propagation framework~\cite{droulias2024ris} to characterize when the physics supports near-field focusing versus far-field beamforming~\cite{droulias2024near}, providing a rigorous reference before introducing detailed site models.
These analytical foundations define the sensitivity and feasibility regions to guide the digital twin simulations.

\noindent\textbf{Stage 2: High-fidelity digital twin simulation.}
Stage 2 increases realism while retaining control and repeatability. It is more complex to execute than Stage 1 but sufficiently scalable for broad design space exploration and sensitivity analysis under controlled variability. In Stage 2, the evaluation becomes operational in controlled settings: definitions of outage, rate of periodic updates, overhead accounting, and success criteria (e.g., meeting specific accuracy or latency thresholds) are fixed and applied consistently across baselines, following Table~\ref{tab:protocol_checklist}. We integrate ray tracing tools, such as Sionna RT~\cite{hoydis2023sionna}, with site-specific map data to generate received power heatmaps and channel impulse responses~\cite{he2019design}. This enables physics-based modeling of multipath propagation for indoor and outdoor scenarios, including reflection, diffraction, and scattering effects that dominate realistic operation \cite{koivunen2024multicarrier}. Stage 2 is also suitable to evaluate robustness against mobility, interference regimes, and impairment profiles while preserving statistical rigor through Monte Carlo (MC) runs and confidence intervals.
Verifying that these simulated findings hold in practice requires transitioning from the digital twin to physical platforms and over-the-air measurements.

\noindent\textbf{Stage 3: Experimental validation via emulation and over-the-air testbeds.}
Stage~3 provides the highest realism by exposing the closed-loop \ac{AI}-native \ac{ISAC} pipeline to hardware and deployment impairments that are difficult to capture in simulation, including synchronization offsets, calibration drift, front-end nonlinearities and quantization, and mobility-induced dynamics. To preserve traceability while increasing realism, Stage~3 follows a reproducible protocol that combines (i) trace-based replay/emulation and (ii) live over-the-air operation on the target platform. Trace-based replay logs the full observation--decision--action loop (sensing observables, inferred state, control actions, timestamps, and configuration), enabling controlled ablations and impairment injection offline, whereas over-the-air trials validate performance in real time under the true sensing/communication/control feedback loop.

%% file: scenarios.tex
\section{Application Scenarios}
To demonstrate how the proposed methodology may be applied in practice, we illustrate it in two representative scenarios sharing a common objective: coverage extension for sensing-enabled services under practical constraints (e.g., limited energy budgets, restricted pilot overhead, RIS size, and strict latency requirements). The scenarios differ primarily in the actuation mechanism used to extend coverage: (i) \ac{UAV}-based mobility control for outdoor coverage and (ii) \ac{RIS}-based environment programmability for indoor coverage. In both cases, we evaluate the system as an \ac{AI}-native closed-loop \ac{ISAC} pipeline where sensing performance, communication reliability, and control overhead are intertwined. Beyond illustrating the evaluation workflow, we translate these high-level outcomes into measurable metrics through explicit rules and reporting conventions following Table~\ref{tab:protocol_checklist}. These include outage definitions, accuracy thresholds, baseline comparisons, and distribution-based reporting (CDFs or percentiles) with multiple trial repetitions and statistical uncertainty margins.

\subsection{Use case 1: \ac{UAV}-based Outdoor \ac{ISAC} Coverage Extension}
The outdoor scenario targets extending coverage over large or obstructed areas, such as urban canyons or disaster zones, using \ac{UAV}-mounted access points that serve as mobile sensing and communication nodes. Coverage extension is achieved primarily through trajectory control. 
The \ac{UAV} repositions and adjusts its height to extend the radar horizon and improve \ac{LoS} probability and observation geometry while respecting flight, energy, and update rate constraints.
The evaluation emphasizes positioning availability and localization accuracy, coupled with mission-level constraints such as endurance, update periodicity, and communication overhead, following the operational definitions in Table~\ref{tab:protocol_checklist}. A corresponding baseline is a fixed or pre-planned trajectory, e.g., a lawnmower or circular scan, which isolates the value of closed-loop mobility control under matched probing and signaling overhead.

In our instantiation, \ac{UAV}-supported sensing-enabled coverage extension is illustrated by combining improved platform self-localization via 5G and \ac{GNSS} fusion~\cite{campolo20245gnss} with victim localization and search and rescue workflows~\cite{blaga20243dsar}. Consistent with the proposed methodology, we utilize the following procedure: i) use Stage 1 to screen feasibility and sensitivity regarding observation geometry and \ac{LoS} assumptions; ii) use Stage 2 to quantify end-to-end trade-offs in a digital twin, reporting distributions of localization error alongside latency and overhead. In Section~\ref{sec:results}, we add an illustrative multi-stage example for the outdoor \ac{UAV} scenario combining Stage 1 and Stage 2 results with preliminary Stage 3 over-the-air \ac{UAV} validation.

\subsection{Use case 2: \ac{RIS}-based Indoor \ac{ISAC} Coverage Extension}
The indoor scenario targets coverage extension in reflective and cluttered environments such as industrial spaces, where \ac{GNSS} is unavailable and \ac{NLoS} is frequent. \acp{RIS} reshape propagation and provide additional controllable paths that improve both connectivity and radio-based sensing in occluded areas. The evaluation focuses on positioning availability and localization accuracy as primary sensing outcomes, complemented by joint metrics such as end-to-end latency and sensing and control overhead, all defined and reported using the operational rules of Table~\ref{tab:protocol_checklist}.

To quantify coverage extension, we report spatial \ac{RSRP} maps and CDFs, defining outage as \ac{RSRP} below $-100$~dBm. Coverage extension is measured as the relative increase in area, or fraction of locations, meeting the KPI thresholds compared to a baseline without \ac{RIS} actuation or with static \ac{RIS} settings, isolating the value of environment programmability. We also consider representative 2~GHz, 3.5~GHz, and 10~GHz settings to capture frequency-dependent multipath and blockage effects under common geometry and reporting rules. Localization ground truth is provided by a LiDAR-based reference, giving confidence-bounded error distributions.

%% file: results.tex
\section{Illustrative Multi-Stage Assessments}
\label{sec:results}

\subsection{UAV-based Outdoor ISAC Coverage Extension}

This section illustrates how the proposed methodology is applied for the outdoor \ac{UAV} coverage extension scenario. The goal is not a comprehensive performance study, but a reproducible pattern linking Stage 1 feasibility screening, Stage 2 digital twin evaluation, and Stage 3 experimental confirmation under consistent assumptions, baselines, and KPI to KVI mappings (Table~\ref{tab:protocol_checklist}).

\begin{figure*}[!t]
\centering
\subfigure[Stage 1: sensitivity screening (adapted from~\cite{blaga20243dsar})]{
\includegraphics[width=0.40\textwidth]{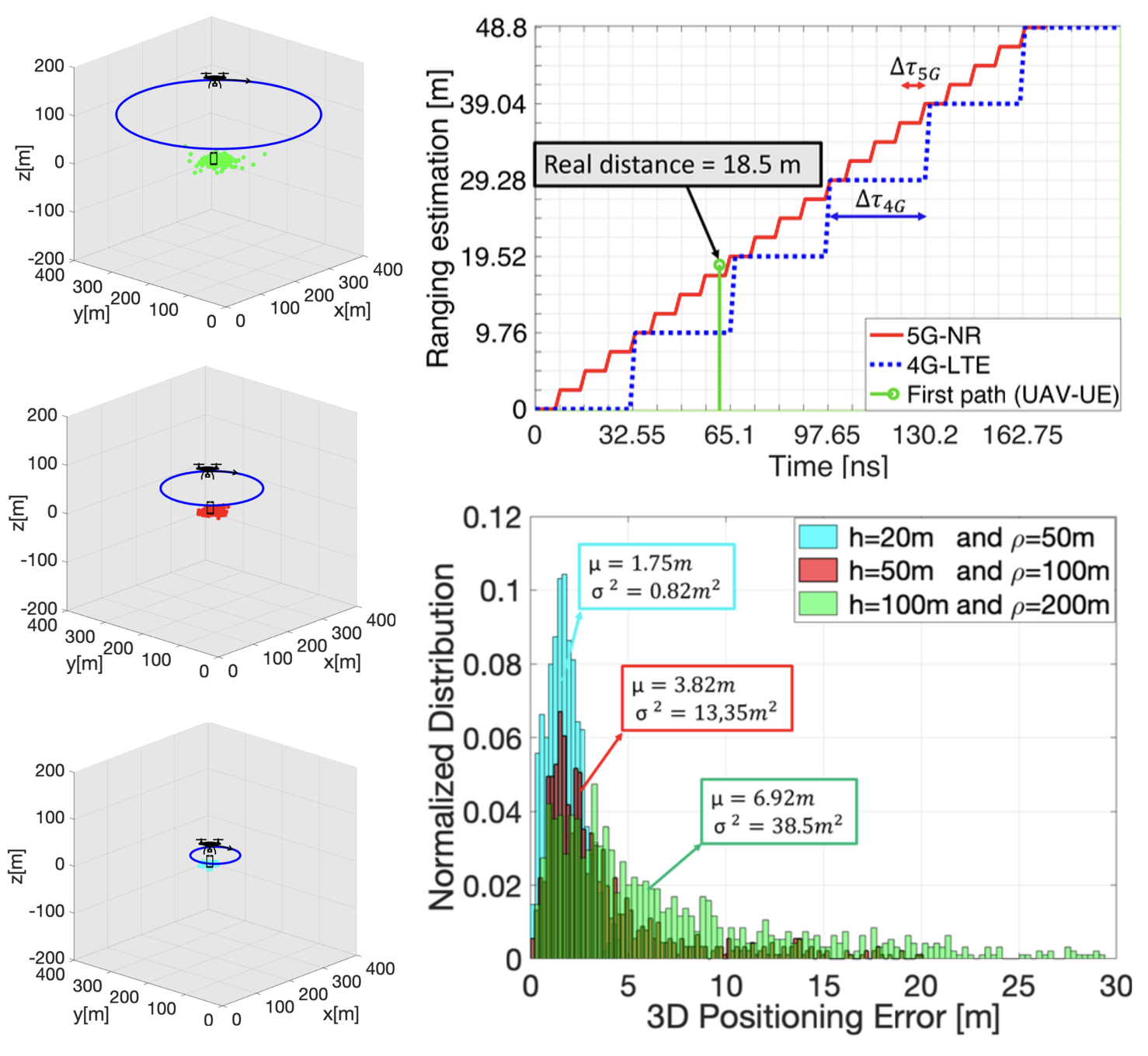}}\hfill
\subfigure[Stage 2: digital twin trade offs and distributions (adapted from~\cite{blaga20243dsar})]{
\includegraphics[width=0.57\textwidth]{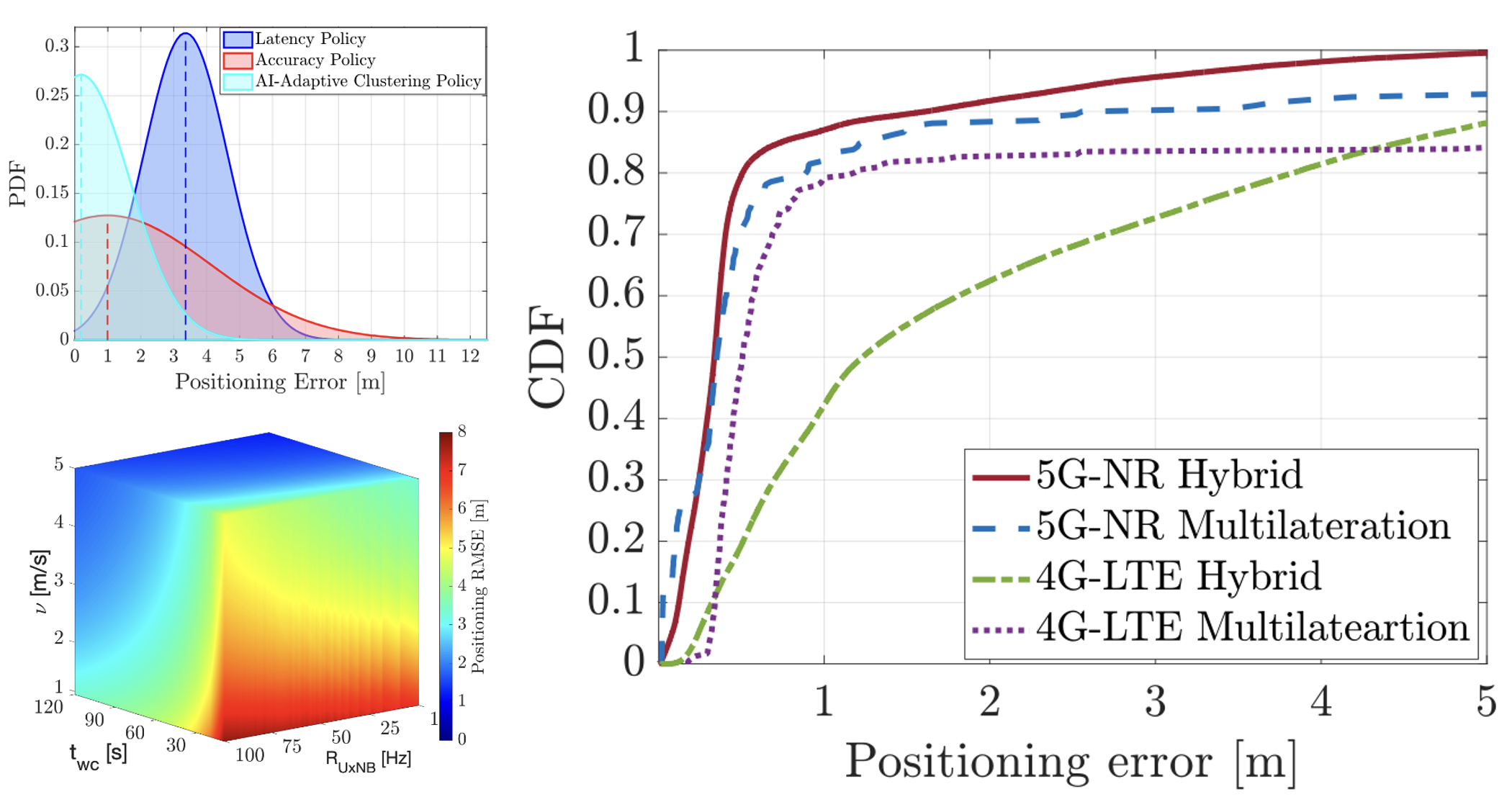}}\vfill
\subfigure[Stage 3: over-the-air validation in DroneLab]{
\includegraphics[width=0.86\textwidth]{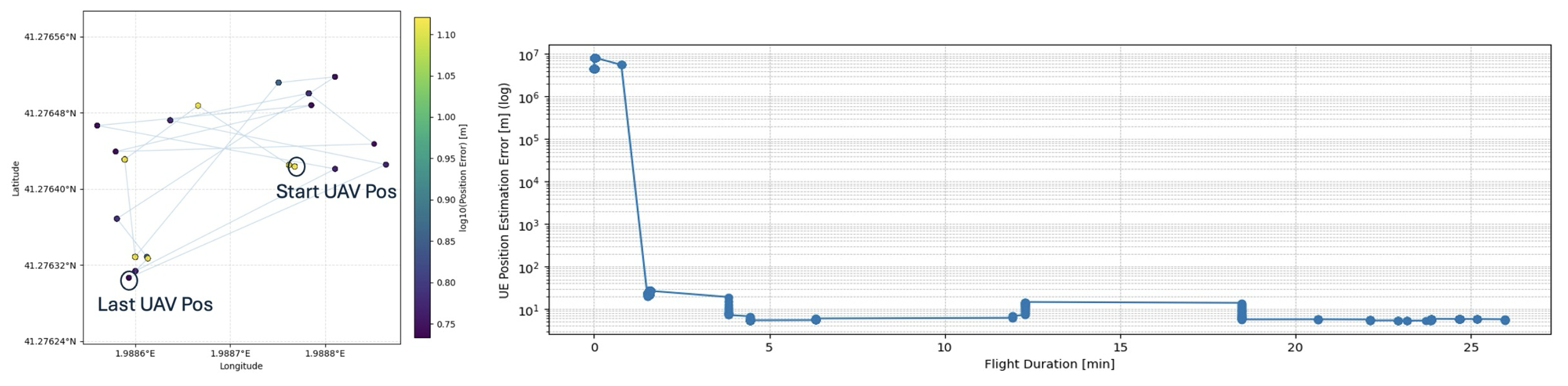}}
\vspace{-1mm}
\caption{Illustrative multi-stage example for the outdoor \ac{UAV} scenario. Stage 1 screens feasibility and sensitivities to set sweep ranges and baselines. Stage 2 reports distributions and trade offs under matched overhead and fixed operational rules. Stage 3 reports an over-the-air validation in the DroneLab testbed, illustrating the end-to-end pipeline under real-world impairments (mobility, intermittency, interference) and the associated robustness and logging requirements.}
\label{fig:uav_multistage_example}
\vspace{-3mm}
\end{figure*}

\noindent\textbf{Stage 1 (Figure~\ref{fig:uav_multistage_example}a)}
Stage 1 screens feasibility and sensitivities under controlled assumptions to select meaningful operating regions for Stage 2 sweeps, including \ac{UAV} altitude, standoff distance, probing periodicity, and baseline choices (e.g., fixed scan trajectory versus adaptive trajectory control). A practical outcome for uplink \ac{SRS}-based ranging is the coarse delay bin spacing induced by sampling frequency and FFT size. In the processing chain, nominal delay bin spacings are approximately $\Delta \approx 4.9$~m for a 4G LTE configuration and $\Delta \approx 2.4$~m for a 5G NR configuration, setting the coarse range resolution prior to sub-sample refinement.
A simple two-path illustration highlights the impact of temporal resolution on \ac{ToF} estimation: for a true first path of $18.5$~m (with a reflected path of $21$~m), the coarse estimator yields $9.76$~m (LTE) and $17.08$~m (NR).
These results motivate joint consideration of the effects of bandwidth, waveform configuration, and multipaths in Stage 2, and justify hybrid sensing and policy design rather than ranging \ac{ToF}-only.

\noindent\textbf{Stage 2 (Figure~\ref{fig:uav_multistage_example}b)}
Stage 2 evaluates end-to-end behavior in a high-fidelity digital twin with controlled variability, while fixing operational rules and matching overhead across baselines following Table~\ref{tab:protocol_checklist}. In a representative comparison, \ac{ToF}-only multilateration in 4G LTE yields positioning errors exceeding $5$~m in over $20\%$ of cases; 5G NR improves performance but still reaches up to about $2.5$~m at the $90$th percentile.
A hybrid approach combining \ac{ToF} and \ac{AoA} reduces errors, with over $90\%$ below $1.5$~m in 5G NR.
Policy-level evaluation shows the accuracy--latency trade off under matched resource use: a latency-oriented policy yields a median error of $3.36$~m with standard deviation $1.27$~m; an accuracy-oriented policy yields $0.99$~m median error with standard deviation $3.13$~m but in higher runtime, while an adaptive clustering policy reaches $0.19$~m median error with standard deviation $1.47$~m and with best overall accuracy at higher estimation duration.
Finally, trajectory prediction further reduces errors from about $7$~m to $1.1$~m (median) and from $15$~m to $1.9$~m ($80$th percentile).

\noindent\textbf{Stage 3 (Figure~\ref{fig:uav_multistage_example}c)}
Stage 3 validates end-to-end operability under real-world impairments using an over-the-air \ac{UAV} campaign in the DroneLab testbed (Castelldefels, Spain). The setup uses an OpenAirInterface (OAI) \ac{5G}-NR configuration in band~n78 with 40~MHz bandwidth and a USRP~B210 (data captured during \ac{UAV} flight). The platform is a DJI Matrice~350~RTK executing a circular trajectory (left panel), with a flight duration of approximately 25~minutes (middle panel). \ac{UE} positioning is obtained via \ac{ToF} based multilateration fused with the \ac{UAV} RTK positions.
A key observation is intermittent degradation: the elevated-error interval (12--18~minutes) indicates transient interference or unreliable UE--gNB link conditions, producing corrupted \ac{ToF} samples and failures not captured in simulations. This motivates Stage-3 reporting beyond mean accuracy, including robustness-to-impairments, failure-mode characterization, and confidence-aware summaries, supported by full system logging for trace replay and ablations as defined in Table~\ref{tab:protocol_checklist}.

\begin{figure*}[!t]
\centering
\subfigure[Stage 1: feasibility analysis using \ac{FI} (adapted from \cite{encinas2025coloris})]{
\includegraphics[width=0.38\textwidth]{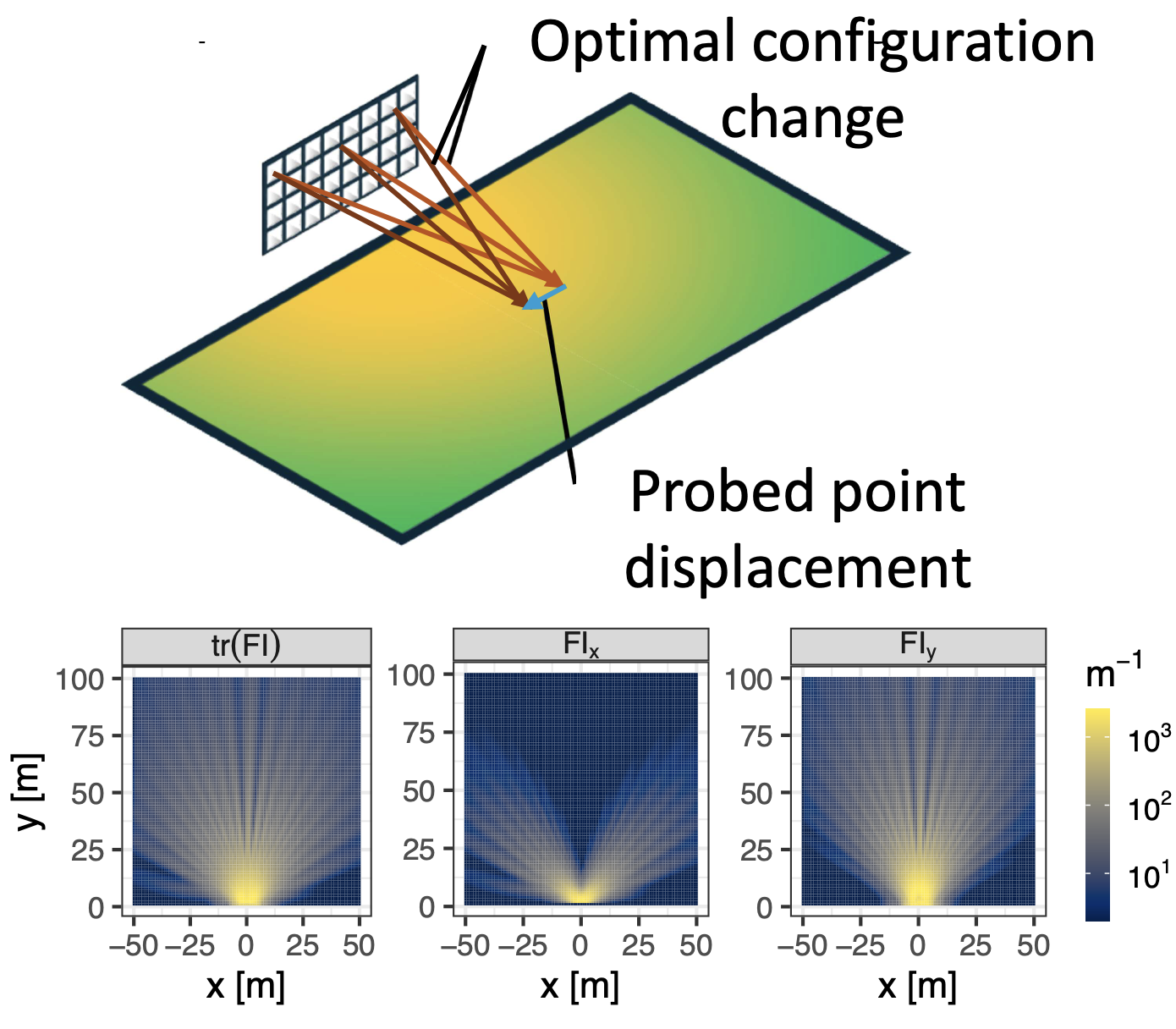}}
\subfigure[Stage 2: digital twin simulation (adapted from \cite{encinas2025coloris})]{
\includegraphics[width=0.42\textwidth]{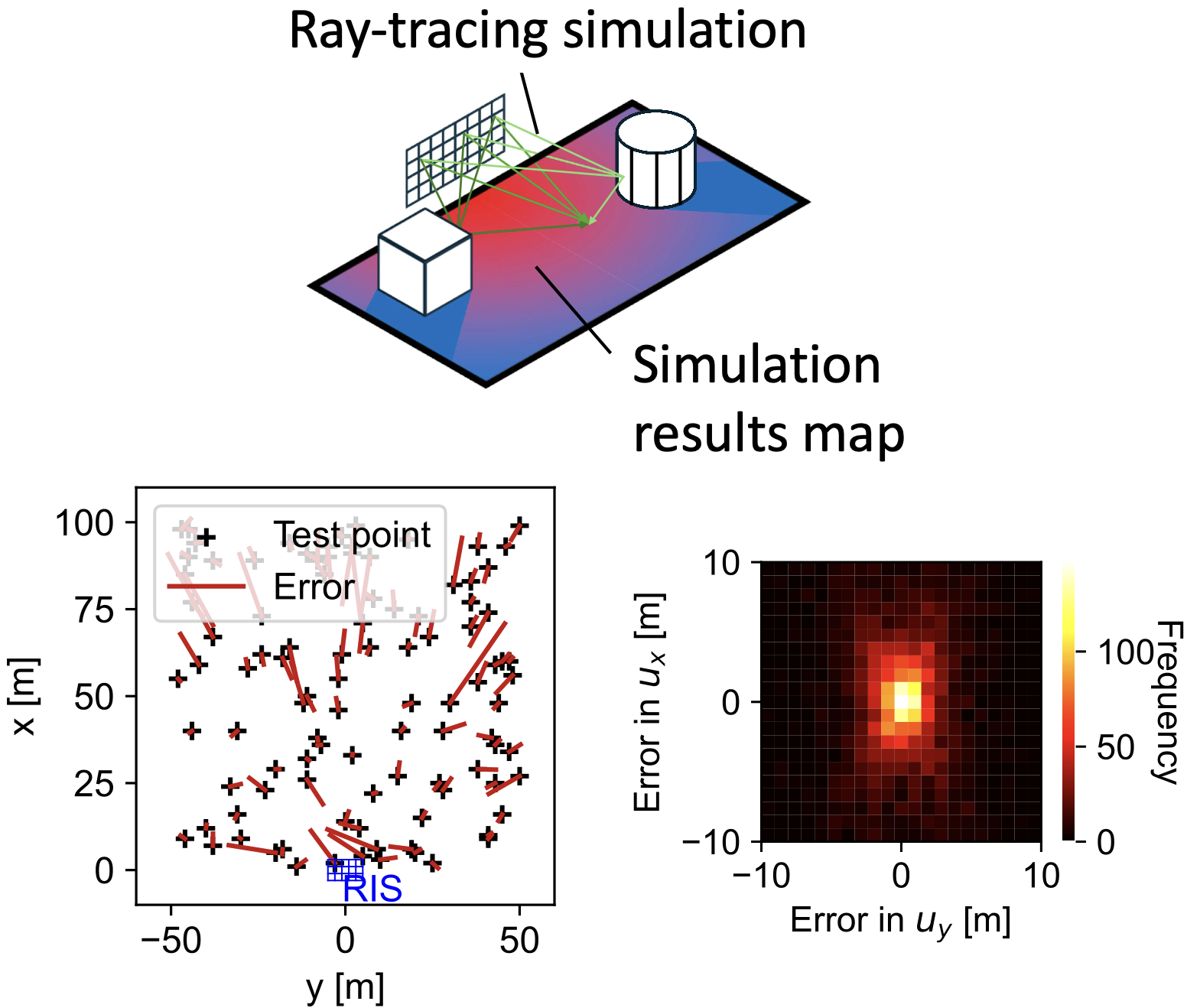}}
\subfigure[Stage 3: experimental validation]{
\includegraphics[width=0.72\textwidth]{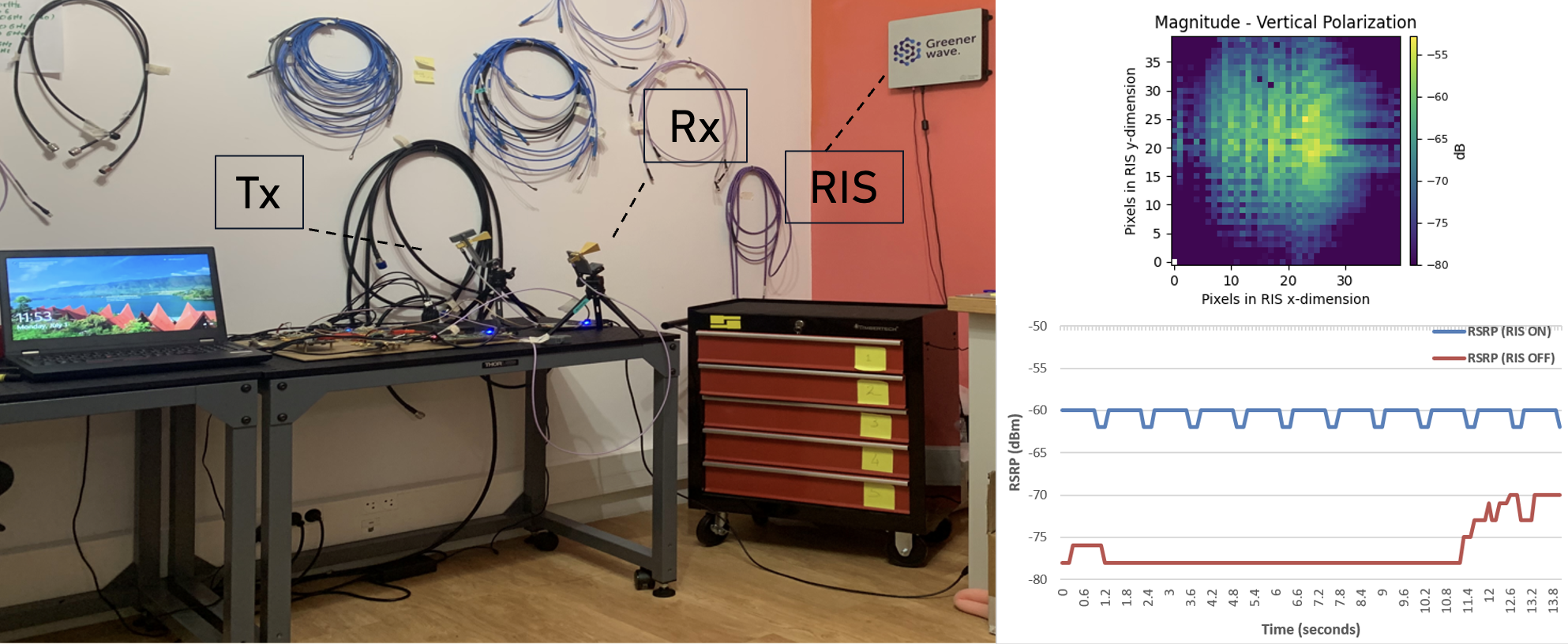}} 
\vspace{-1mm}
\caption{Illustrative multi-stage example for the indoor \ac{RIS} coverage extension scenario. Stage 1 demonstrates feasibility through \ac{FI}. Stage 2 shows performance in the presence of obstacles and complex propagation. Stage 3 confirms the system operation in laboratory conditions.}
\vspace{-3mm}
\label{fig:ris_multistage_example}
\end{figure*}

\vspace{-1mm}
\subsection{RIS-supported Indoor Coverage Extension}
We next illustrate the methodology through a RIS-assisted indoor coverage extension scenario following the COLORIS \cite{encinas2025coloris} framework (for Stages 1 and 2), where RIS configurations optimized for communication are opportunistically reused to support localization and sensing services without dedicated waveforms or localization-specific probing. This example complements the UAV-based outdoor case by illustrating environment programmability as the actuation mechanism enabling closed-loop AI-native ISAC operation.

The considered scenario targets cluttered indoor environments with frequent blockage and NLoS conditions, such as industrial facilities, warehouses, or large indoor venues. A programmable RIS is deployed to create controllable reflected paths that improve both communication coverage and sensing capabilities where direct propagation is weak or unavailable. Unlike conventional RIS-assisted localization approaches requiring dedicated localization phases, configuration sweeps, or additional infrastructure, the considered approach leverages communication-driven RIS configurations as implicit sensing signatures, enabling opportunistic localization with limited additional overhead.

\textbf{Stage 1 (Figure 4a):} Following the three-stage methodology, Stage 1 evaluates feasibility and sensing capability under controlled assumptions. In COLORIS, \ac{FI} and CRB analysis quantify whether RIS configurations carry sufficient spatial information for localization and sensing. The analysis highlights dependence on RIS geometry, phase quantization, operation assumptions, and SNR, and exposes the importance of near-field propagation effects, where RIS configurations depend on angular information, wavefront curvature, and UE distance. The analysis points to a CRB below 1~m in most of the analyzed area, and the dependence of the CRB figure itself on other experimental factors. Rather than claiming deployment-realistic performance, this stage identifies feasibility regions and sensitivity trends guiding Stage 2.

\textbf{Stage 2 (Figure 4b):} This stage introduces site-specific digital twin evaluation with realistic indoor propagation, blockage, reflections, and interference. Following Table I, no-RIS, static-RIS, and adaptive closed-loop RIS baselines are compared under matched signaling and overhead. Localization capabilities are quantified through spatial maps, and the relative area satisfying the adopted KPI thresholds. Localization performance is analyzed through error distributions and percentile-based metrics using map-based ground truth. Within this closed-loop setting, RIS actuation affects both communication performance and sensing observability. By modifying the propagation environment, the RIS improves coverage in occluded regions and increases geometric diversity for localization. As illustrated, the adaptive closed-loop RIS deployment is able to obtain an average localization accuracy of 5~m in spaces of 100~$\times$~100~m$^2$, while benchmark approaches are not able to obtain localization solutions under the same conditions. Low computational complexity architectures operating in representative indoor-like spaces of 10~$\times$~10~m$^2$ are able to locate users within 0.7~m. This illustrates a key aspect of AI-native ISAC evaluation: sensing observability is not only environment-dependent but becomes a controllable variable through programmable propagation.

\noindent\textbf{Stage 3 (Figure~\ref{fig:ris_multistage_example}c)}
This stage validates the end-to-end \ac{RIS}-assisted \ac{ISAC} pipeline under realistic hardware and deployment impairments. The experimental evaluation utilizes a programmable Greenerwave \ac{RIS} alongside dedicated transmitter and receiver antennas to assess environment programmability under operational conditions. The illumination footprint on the \ac{RIS} (captured via vertical polarization magnitude) verifies the effective spatial distribution of incident waves. Furthermore, over-the-air measurements confirm a substantial improvement in connectivity, with Reported \ac{RSRP} demonstrating a gain of approximately 18~dB when the \ac{RIS} actively optimizes reflected paths compared to the static baseline without \ac{RIS} actuation. These results confirm the practical viability and significant coverage extension capabilities of the considered \ac{RIS} prototypes in real-world deployments.

%% file: conclusion.tex

\section{Challenges and Future Directions}
Evaluating \ac{AI}-native \ac{ISAC} as a closed-loop system raises challenges future methodologies must address to produce robust, reproducible deployment evidence. These challenges are especially visible in coverage extension, where actuation (e.g., \ac{RIS} reconfiguration or \ac{UAV} mobility) couples sensing quality, communication reliability, and operational constraints, and where results must transfer across environments and hardware.

\textbf{Energy Footprint of Network Intelligence.}
A key concern is the energy footprint of the intelligence layer. While \ac{RIS} can reduce radiated power by shaping propagation and \ac{UAV} mobility can improve geometry and \ac{LoS} likelihood, sensing, inference, and control incur non-negligible compute and signaling cost. Future evaluation pipelines should therefore report energy per update or decision and attribute it across the end-to-end chain, including radio front-end, baseband \ac{DSP}, \ac{AI} inference, actuation, and signaling. This is essential for battery-constrained platforms such as \ac{UAV}s and \ac{IoT} endpoints, where reduced transmit power may be offset by increased compute, reconfiguration, and coordination overhead \cite{albanese2024ares}. Without this holistic view, apparent physical-layer gains may be negated at the system level, leading to misleading conclusions.

\textbf{Robustness Under Uncertainty and Distribution Shifts.}
A second challenge is robustness of learning-based policies under uncertainty and distribution shift. Many studies assume training conditions represent deployment, yet realistic environments exhibit non-stationary interference, topology changes, mobility, and multipath variation. Methodologies should therefore include stress-testing protocols that quantify resilience under distribution shift using degradation curves, recovery time, and uncertainty-aware performance. Evaluations should also document failure modes and fallback behavior under synchronization loss, calibration drift, and partial hardware faults, reporting whether minimum guarantees (e.g., availability and latency bounds) are preserved. This robustness assessment can be enabled by combining calibrated digital twins with trace replay and controlled impairment injection.

\textbf{Impact of Bandwidth on AI Decisions.}
Another challenge is the impact of physical-layer resource constraints on the network intelligence layer. Specifically, the precision of environmental sensing is strictly bounded by the available channel bandwidth. In limited-bandwidth regimes, the \ac{ISAC} system extracts coarse contextual details, leading to noisy, low-resolution state representations. When \ac{AI} agents are trained or operated on such degraded observations, it limits the model's ability to learn accurate policies, often resulting in flawed actuation or resource-allocation decisions. Future evaluation pipelines must explicitly quantify how varying bandwidth allocations, and the resulting sensing uncertainty, propagate through the \ac{AI} model, stress-testing the closed-loop control against degraded input data.

\textbf{Continuity Across Evaluation Stages.}
A recurring bottleneck is the lack of continuity between analytical, simulation, and experimental stages. To improve transferability, future work should emphasize cross-stage alignment through explicit logging and release of configuration and trace artefacts, including observables, decisions, and actuation commands, seeds, and KPI rules. Stage 2 digital-twin models should be calibrated against measurements when possible, and studies should report sensitivity to key assumptions (e.g., synchronization offsets and update periodicity), not just best-case gains. For \ac{UAV}-based systems, this motivates measurement campaigns capturing uplink reference signals under controlled geometries and mixed \ac{LoS}/\ac{NLoS} conditions using SDR platforms and onboard compute for repeatable benchmarking.

\textbf{Privacy and Security Implications.}
As sensing-capable waveforms become pervasive, privacy and security risks increase, including unintended or unauthorized sensing. Evaluation pipelines should therefore integrate security and privacy indicators alongside classical \ac{ISAC} KPIs, such as detectability of sensing activity, robustness to inference attacks, and user exposure impact. Including these metrics helps prevent performance gains from undermining trust or regulatory compliance.

\section{Conclusion}

This article presented a unified methodology for \ac{AI}-native \ac{ISAC}, enabling a traceable transition from bounds and feasibility analysis to high-fidelity digital-twin simulation and experimental validation. By aligning assumptions, baselines, and KPI/KVI reporting across stages, the framework supports repeatable benchmarking of closed-loop behavior across heterogeneous deployments. The \ac{RIS}-based indoor and \ac{UAV}-based outdoor instantiations illustrate how to operationalize coverage extension and positioning availability through explicit rules, matched-overhead baselines, and distribution-based reporting. More broadly, the proposed methodology serves as a design and verification tool helping translate theoretical \ac{ISAC} potential into deployment-ready performance claims under realistic constraints.